\documentstyle[preprint,aps,psfig]{revtex}

\tightenlines

\begin{document}

\draft

\title{Observation of the east-west anisotropy of the atmospheric neutrino flux}
\date{Jan. 12, 1999}

\maketitle


{\center \large The Super-Kamiokande Collaboration\\}

\begin{center}
\newcounter{foots}

T.Futagami$^u$,
Y.Fukuda$^a$, T.Hayakawa$^a$, K.Inoue$^a$,
K.Ishihara$^a$, H.Ishino$^a$, Y.Itow$^a$,
T.Kajita$^a$, J.Kameda$^a$, S.Kasuga$^a$
% ...
\addtocounter{foots}{1}
\addtocounter{foots}{1}
\addtocounter{foots}{1} 
L.L.Wai$^{w,\fnsymbol{foots}}$, 
R.J.Wilkes$^w$, K.K.Young$^{w,\dagger}$

\footnotesize \it

$^a$Institute for Cosmic Ray Research, University of Tokyo, Tanashi,
Tokyo 188-8502, Japan\\
% ...
$^u$Department of Physics, Tokyo Institute of Technology, Meguro, 
Tokyo 152-8551, Japan \\
% ...
$^w$Department of Physics, University of Washington,    
Seattle, WA 98195-1560, USA    \\
\end{center}

\begin{abstract}

  The east-west anisotropy, caused by the deflection of primary cosmic rays
 in the Earth's magnetic field, is observed for the first time in the
 flux of atmospheric neutrinos. Using a 45~kt$\cdot$year exposure of the
 Super-Kamiokande detector, 552 $e$-like and 633 $\mu$-like 
 horizontally-going
 events are selected in the momentum range between 400 and 3000 MeV/$c$.
 The azimuthal distribution of $e$-like and $\mu$-like events
 agrees with the expectation from atmospheric neutrino flux
 calculations that account for the geomagnetic field,
 verifying that the geomagnetic field effects in the
 production of atmospheric neutrinos in the GeV energy range are
 well understood.
 
\end{abstract}

\begin{figure}[h]
\begin{center}
\psfig{figure=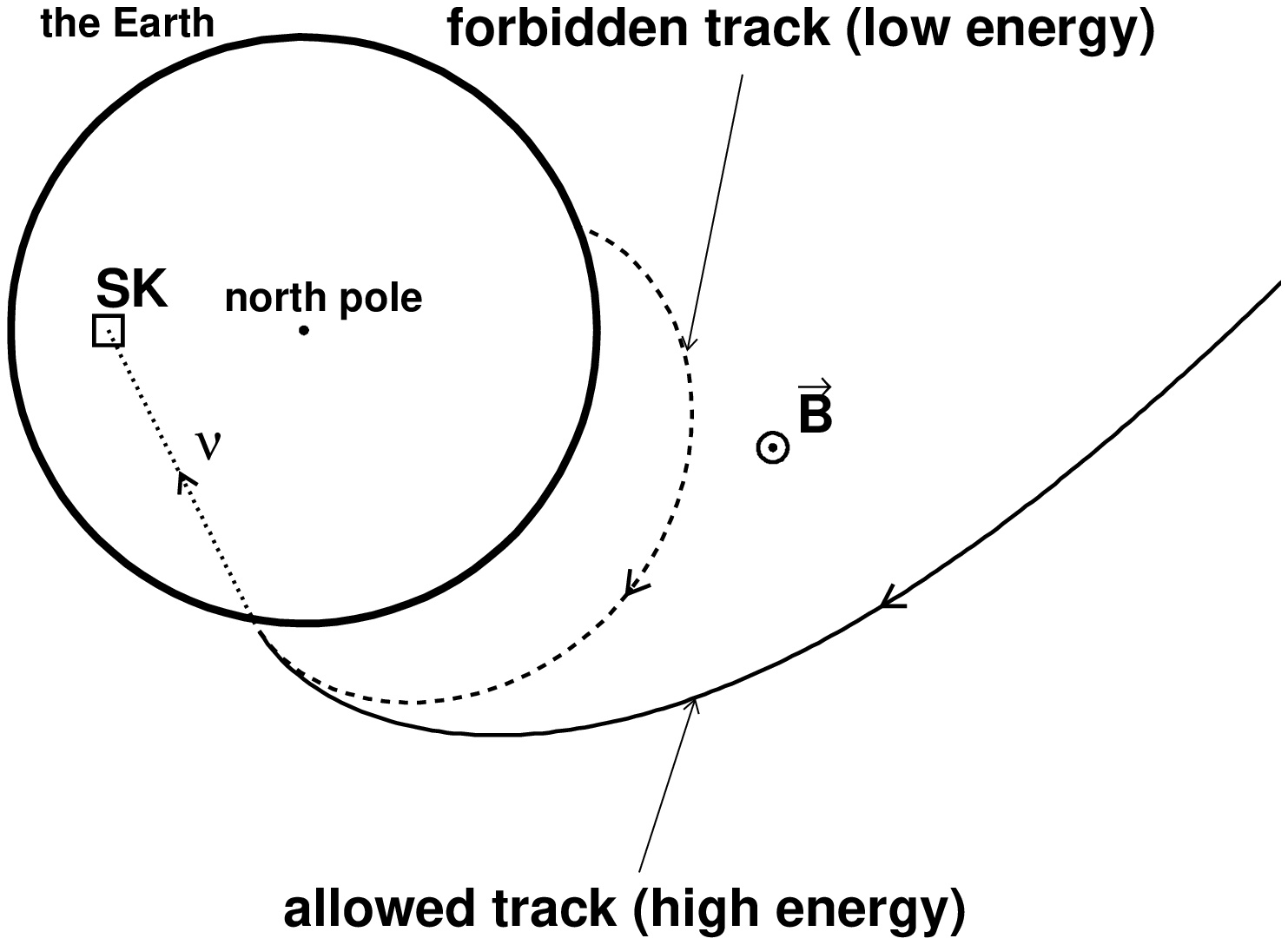,width=9cm}
\caption{Typical allowed and forbidden tracks of the 
incident cosmic rays.
The curved line shows the allowed trajectory of the cosmic ray
which has momentum larger than cut-off momentum and
the dotted curved line shows forbidden trajectory with momentum below 
cut-off. 'SK' represents the Super-Kamiokande detector.}
\label{fig:cutoff} 
\end{center}
\end{figure}

 The primary cosmic ray flux approaching the Earth is known to be almost
isotropic{\cite{Munakata}}.
However, since the Earth has a magnetic field and the primary cosmic rays
are positively charged, an angular anisotropy is produced
for those primaries which reach and interact in the atmosphere. 
 This azimuthal anisotropy, called the east-west effect,
was discovered in the 1930's as a deficit
of secondary cosmic rays (muons) arriving from the easterly
direction compared to a westerly direction
{\cite{CosmicMu1}}. In fact, this effect was used to
infer that the primary cosmic rays are positively charged.
The azimuthal anisotropy varies according to
the position on the Earth and the momentum and
charge of the primary cosmic ray.
 The anisotropy is largest for the lowest momenta particles.  
 The anisotropy can be characterized by a cut-off momentum, 
the maximum momentum at which the incoming particle cannot reach the 
atmosphere. 
At the Super-Kamiokande experiment, the cut-off momentum for protons averages
about 10~GeV/$c$. Since Super-Kamiokande is located close to the geomagnetic
equator (25.8$^{\circ}$~N geomagnetic latitude),
 the cut-off momentum for horizontally arriving protons from the east is
$\sim$50~GeV/$c$, considerably higher than the average over directions.
Figure \ref{fig:cutoff} illustrates typical allowed and 
forbidden tracks of cosmic rays. 
This results in the depletion of primary cosmic ray interactions east 
of Kamioka. Neutrinos produced in the atmosphere by these cosmic
rays should likewise be depleted in the easterly direction. 

 In this paper
we report the first observation of the east-west anisotropy in the
atmospheric neutrino flux. We observe a neutrino east-west effect 
which is in agreement with expectations based on detailed flux calculations
and a Monte Carlo simulation of neutrino interactions.
This confirms that the angular dependance of the atmospheric neutrino
flux calculations in the GeV energy range are reasonable.

 The Super-Kamiokande detector is a 50 kt water Cherenkov detector
located at the Kamioka Observatory, ICRR, Univ. of Tokyo.
The center of the detector is at $36^{\circ} 25' 33''$ N, $137^{\circ} 
18' 37''$ E, 371.8~m above the sea level.
 To reduce cosmic ray muons, it is placed at 2700 meters-water-equivalent
 below the peak of Mt. Ikenoyama. It is cylindrically shaped with the height 
of 41.4 meters and the diameter of 39.3 meters, and divided into the 
inner and outer detectors by a stainless steel support structure
with a pair of opaque sheets.
The inner detector is viewed by 11,146 photo-multiplier tubes (PMTs) of 
50-cm diameter, facing inward to detect the neutrino
events. The outer detector, a layer of water of 2.6 to 2.75 m thick, 
has 1885 PMTs of 20-cm diameter used to veto entering events and 
to tag exiting events.

 The energy and direction of a neutrino are
measured from the momentum of the final state charged lepton 
which is produced in a charged current interaction of the neutrino 
on nucleus, $\nu + {\rm N} \to l + X $. The flavor of the final state 
lepton is used for tagging the flavor of the neutrino.
 With the 45~kt$\cdot$yr exposure of the Super-Kamiokande,
a total of 4077 fully-contained (FC) single-ring atmospheric neutrino events
were observed with the following requirements: (1) total number of hits
in outer detector less than 25, and no spatial cluster with more than
10 hits, (2) total charge collected in the inner detector more 
than 200 p.e.s, (3) the ratio (maximum p.e. in any single PMT)/(total p.e.s)
less than 0.5, (4) the time interval from preceding event larger than 
100 $\mu s$, (5) the vertex position should be inside the fiducial volume,
 2 meters inside from the inner detector wall, and 
(6)single-ring events with the momentum higher than 100 MeV/$c$
and 200 MeV/$c$ for $e$-like and $\mu$-like events, respectively.
To determine the direction of neutrinos from the observed
leptons, only single Cherenkov ring events were used.
 Then they were separated into $e$-like and $\mu$-like events 
according to their showering or non-showering signature. 
The particle identification efficiency was estimated to be better than
99\% {\cite {SubGeV}}.

For analysis of the east-west anisotropy, the following additional 
cuts were applied:
 (7) momentum was required to be between 400 MeV/$c$ and 3000 MeV/$c$ 
for both $e$-like and $\mu$-like events
 and
 (8) cosine of the zenith angle of the leptons was required to be between
 -0.5 and 0.5 .
Criterion (7) selects the momentum region where the east-west anisotropy is
expected to be most significant.
The east-west anisotropy of the flux is expected to be larger for 
lower energy neutrinos
because of the cutoff momentum. On the other hand, the poor determination
of the neutrino direction at low energy tends to wash out the flux 
anisotropy. The estimated mean scattering angle in the momentum region of 
criterion (7) is 36 degrees for both $\nu_{\rm e}, \nu_{\mu}$ charged current
events.
The contamination of neutral current events in these $e$-like and $\mu$-like
events were estimated to be 9.0\% and 1.3\%, respectively.
Criterion (8) is to select neutrino events from near the horizon 
where the effect of the geomagnetic cutoff is maximum.

\begin{figure}[h]
\begin{center}
\psfig{figure=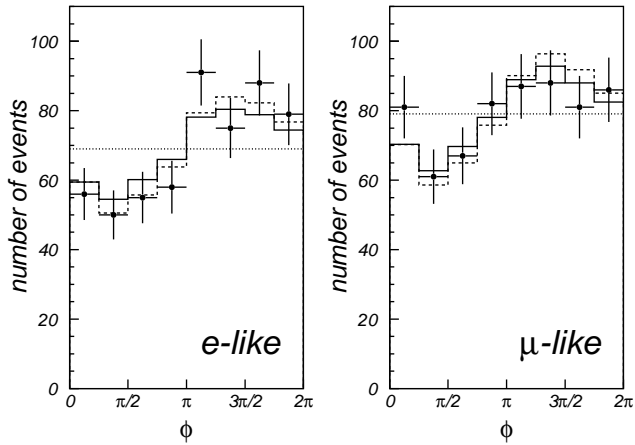,width=9cm}
\caption{Azimuthal angle distributions of $e$-like and $\mu$-like events.
The crosses
represent the data points, the histogram drawn by solid line (dashed-line)
shows the prediction of the Monte Carlo based on the flux of Ref.
\protect\cite{Honda} (\protect\cite{Bartol}).
Data are shown with statistical errors. The Monte Carlo has 10 times more
statistic than data. The Monte Carlo histogram is normalized to the total
number of the real data.
$\phi$ represents the azimuthal angle. $\phi$ = 0, $\pi/2$,
$\pi$ and $3\pi/2$ shows particles going to north, west, south, and east,
respectively.}
\label{fig:chi2} 
\end{center}
\end{figure}

 After these criteria, 552 $e$-like and 633 $\mu$-like events remained.
These data were compared with Monte Carlo simulations with two independent
flux calculations{\cite {Honda,Bartol}}. These calculations
represented the geomagnetic field by a multipole expansion of the
spherical harmonic function.
Cutoffs are calculated for all zenith and azimuthal angles at
the Super-Kamiokande detector site by backtracing antiprotons through 
the three dimensional map of the geomagnetic field.
 Figure \ref{fig:chi2} shows the azimuthal
 distributions for $e$-like and $\mu$-like events.
The number of Monte Carlo events was normalized to the number of data
events to compare the azimuthal shape without regard to the 20\%
uncertainty in normalization and the deficit of $\mu$-like events
most likely due to neutrino oscillations {\cite {SubGeV,MultiGeV,neuosc}}. 
The Monte Carlo events we used here assumed no neutrino oscillations.
 The lengths of the
flight paths of the neutrinos vary with zenith angle but do not 
vary with the azimuthal angle.  Therefore, neutrino flavor oscillations 
have little effect on east-west anisotropy. The estimated effect 
is far below the statistical accuracy of the data.

The azimuthal angles are divided into 8 bins, 45 degrees each.
Agreement between the data and the Monte Carlo is good: the 
$\chi^2$ values were 5.1/7DOF and 2.6/7DOF for $e$-like and
$\mu$-like events, respectively.

 A quantitative comparison between the data and Monte Carlo
was performed using a Kuiper test {\cite {stephens70,NumRecipesF}}.
This test calculates a binning and starting-point free probability
that observed data are the result of an assumed distribution.
The Kuiper statistic $V$ is defined as:
\[
\displaystyle V = \max_{0<\phi<2\pi} [S_N(\phi)-P(\phi)]+
\max_{0<\phi<2\pi} [P(\phi)-S_N(\phi)],
\]
where $\phi$ is the azimuthal angle,
$S_{N}(\phi)$ is a cumulative probability function from data
and $P(\phi)$ is the one from Monte Carlo.
 The significance is obtained from the statistic
$V^{*}=V(\sqrt{n}+0.155+0.24/\sqrt{n})$, and defined as
\[
\displaystyle Prob=
2 \sum_{j=1}^{\infty} (4j^2V^{*2}-1)\exp(-2j^{2}V^{*2}),
\]
 where $n$ is number of events.

 From this test, the probability that the azimuthal distribution of the data
originated from a flat parent distribution
was 0.0008\% (20\%) for $e$-like ($\mu$-like) events.
The azimuthal distribution of $e$-like events is
inconsistent with a flat distribution at more than 99\% CL.
Also, the probabilities that the data match the Monte Carlo in shape with
the flux of Ref.{\cite{Honda}} were
42\% for $e$-like events and 92\% for $\mu$-like events. 
For a Monte Carlo with neutrino oscillations with
($\Delta m^2$, sin$^2 2\theta$) = (2.2$\times$10$^{-3}$ eV$^2$,1.0)
{\cite {neuosc}}, the probability was the same within 1\% for both
 $e$-like and $\mu$-like events. As expected, the
probability did not change much with the addition of neutrino oscillations.

 With current statistics, the deficit of the westward-going events
is apparent in the $e$-like but not in the $\mu$-like data sample.
The smaller anisotropy observed in $\mu$-like data is
consistent with a statistical fluctuation.  The expected anisotropy
is similar for $\nu_e$ and $\nu_{\mu}$ {\cite {Honda,Bartol}}; 
the calculated anisotropy for $\nu_e$ is only about 1.1~times larger
than that for $\nu_{\mu}$ in the relevant energy region.

\begin{figure}[h]
\begin{center}
\psfig{figure=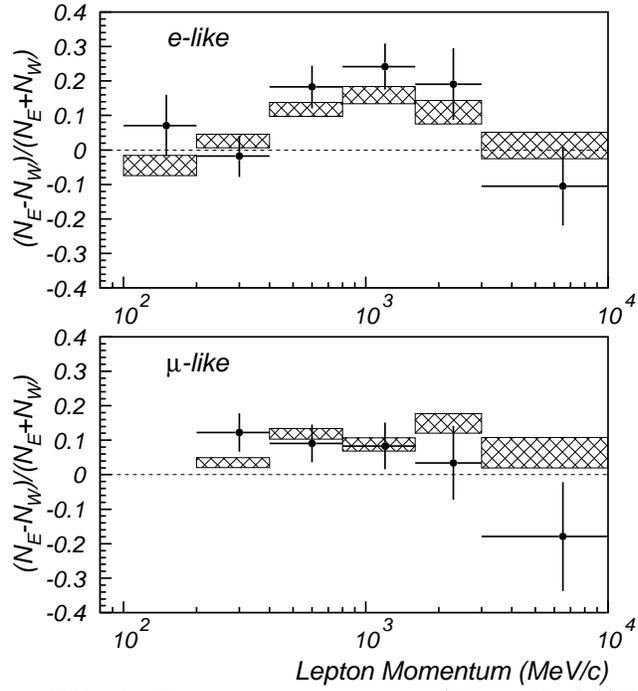,width=9cm} 
\caption{East-west asymmetry $(N_E-N_W)/(N_E+N_W)$ as a function of
lepton momentum for $e$-like and $\mu$-like events.
The crosses with error bars represent the data and the hatched region 
represents the prediction based on the flux of Ref. \protect\cite{Honda}
; error bars are statistical.}
\label{fig:ew_ratio} 
\end{center}
\end{figure}

 The study of the east-west anisotropy as a function of lepton momentum
is important for our understanding of the neutrino-lepton angular
correlation and the geomagnetic field effect on the 
production of atmospheric neutrinos. Figure \ref{fig:ew_ratio} shows
the east-west
asymmetry $(N_E-N_W)/(N_E+N_W)$ as a function of lepton momentum, where
$N_E(N_W)$ represents the number of eastward(westward)-going events.
Here the azimuthal distribution was divided into two bins, and the
same zenith angle cut (8) was used. The data and the Monte 
Carlo agreed well;
the $\chi^2$ values were
5.9/6DOF (5.8/5DOF) for $e$-like ($\mu$-like) events.
The $\chi^2$ values of a comparison of the data and a straight
line at $(N_E-N_W)/(N_E+N_W)$ = 0 were 26.5/6DOF (10.5/5DOF)
for $e$-like ($\mu$-like) events.
As expected from the large neutrino-lepton scattering 
angle, the measured east-west asymmetry was small below 400MeV/$c$. 
Above 3 GeV/{\it c}, the neutrinos originate from primary cosmic rays
that are minimally affected by the geomagnetic field, so the flux
is symmetric from the east and west.

 In summary, we have observed for the first time clear 
evidence for the east-west anisotropy in the atmospheric neutrinos,
based on the data from 45 kt-yr of exposure of the Super-Kamiokande
detector.
A deficit of westward-going $e$-like events
was seen at more than 99\% confidence level,
while for $\mu$-like events, it was statistically less significant.
However, the azimuthal distributions of these events and the
momentum dependence of the east-west anisotropy for both
$e$-like and $\mu$-like agreed well with the
prediction based on detailed neutrino flux calculations.
This observation strongly suggests that the geomagnetic field 
effects in the production of atmospheric neutrinos in the GeV 
energy range are well understood.

 We gratefully acknowledge the cooperation of the Kamioka Mining and Smelting
Company.  The Superk-Kamiokande experiment was built from, and has been 
operated with, funding by the Japanese Ministry of Education, Science,
Sports and Culture, and the United States Department of Energy.



%
%

\unitlength=1mm

%
%

\end{document}